\documentclass[]{spie}  

\usepackage{amsmath,amsfonts,amssymb}
\usepackage{graphicx}
\usepackage[colorlinks=true, allcolors=blue]{hyperref}

\title{A Hybrid R--Theta Fiber Positioning Robot Using a Piezoelectric Bender for Radial Motion}

\author[a]{Michael Schubnell}
\author[a,b]{Tim Fanning}
\author[a]{Andrew Hope}
\author[c]{Joseph Silber}
\author[a]{Greg Tarl\'e}
\author[c]{Nicholas~Wenner}
\affil[a]{University of Michigan, Ann Arbor, MI, USA}
\affil[b]{Indiana University, Bloomington, IN, USA}
\affil[c]{Lawrence Berkeley National Laboratory, Berkeley, CA, USA}

\authorinfo{Further author information:  schubnel@umich.edu}

\begin{document} 
\maketitle

\begin{abstract}
Future massively multiplexed spectroscopic surveys require focal planes with increasingly dense arrays of robotic fiber positioners. The DESI (Dark Energy Spectroscopic Instrument) fiber positioners, based on a two-axis $\theta$--$\phi$ motorized architecture, have demonstrated excellent reliability and performance during operations and provide the benchmark for evaluating new concepts. However, proposed next-generation instruments such as Spec-S5 require a substantially smaller positioner pitch than DESI, making direct scaling of the existing design challenging. Here we describe a hybrid R--$\theta$ fiber positioning concept in which the rotation about the central axis is provided by a DESI-like $\theta$ motor, while the radial displacement is produced by a piezoelectric bimorph bender coupled to a lightweight carbon-fiber spine. This design builds on the heritage of the  the proven central rotation mechanism while replacing the $\phi$ arm with a compact piezoelectric actuator. We discuss the trade-offs between motorized positioners, piezoelectric slip-stick tilting spines, tubular piezo actuators, and direct piezo-bender actuation. A prototype based on a DESI positioner was fabricated and tested using a commercial bimorph bender and an 8-inch carbon-fiber extension. With closed-loop control using feedback from a fiber-view camera the fiber was brought on target after two corrections with residual stability of approximately 2\,$\mu$m RMS. These results indicate that direct piezo-bender actuation is a promising approach for compact fiber positioning, provided that closed-loop operation is addressed. 
\end{abstract}


\section{INTRODUCTION AND MOTIVATION}
\label{sec:intro}  

Wide-field spectroscopic surveys are among the primary tools for studying the origin, structure, composition, and evolution of the universe. Measurements of the three-dimensional distribution of galaxies and quasars provide constraints on dark energy, inflation, the growth of structure, and the gravitational effects of dark matter. The next generation of cosmological and astrophysical surveys will require spectroscopy of tens to hundreds of millions of astronomical sources, motivating instruments with very high multiplexing capability and dense focal-plane fiber arrays.

The Dark Energy Spectroscopic Instrument, DESI, represents the current state of the art in massively multiplexed robotic fiber positioning\cite{Silber23}. DESI employs 5000 robotic fiber positioners mounted at the prime focus of the Mayall 4-m telescope and has demonstrated excellent performance through survey operations. The $\theta$--$\phi$ architecture, based on two miniature brushless DC gear motors per positioner, is field proven, robust, and reliable. As such, DESI provides the baseline against which new fiber-positioning concepts should be compared.

Future massively multiplexed instruments, such as the proposed Spec-S5 facility\cite{spec-s5}, require a significantly higher fiber density than currently realized in DESI. The Spec-S5 reference focal-plane concept, for instance, calls for a positioner pitch of approximately 6.2\,mm, compared to the 10.4\,mm pitch of DESI. This reduction in pitch is challenging for a simple scaled version of the DESI positioner. This is largely because the motor diameter and mechanical packaging limit the extent to which adjacent positioners can be placed in close proximity. Designs such as the Trillium concept\cite{trillium} address this issue through more compact motor integration and transfer gearing, while tilting-spine concepts such as Echidna\cite{Echidna1} naturally lend themselves to small pitch. However, each design involves trade-offs in maturity, complexity, positioning accuracy, control strategy, manufacturability, and optical performance.

In this paper, we describe a hybrid R--$\theta$ fiber positioning robot concept intended to explore an intermediate point in this design space. The design retains a DESI-like central $\theta$ motor for azimuthal actuation, while replacing the $\phi$ arm with a piezoelectric bender that provides radial deflection. The goal is to preserve the advantages of the proven DESI rotation stage, for which rotary motion is a natural match, while investigating whether a compact piezoelectric element can provide reliable and precise radial positioning for dense focal-plane applications.

\section{FIBER POSITIONER DESIGN TRADE-OFFS }
\label{sec:trade}

\subsection{DESI Theta--Phi Positioners - the Baseline}

The DESI positioner uses two rotary axes, referred to as $\theta$ and $\phi$, to move the fiber tip within a patrol region. This architecture has several important advantages: it is mechanically well understood, provides deterministic motion, supports stable fiber placement, and has demonstrated long-term operational reliability in a large astronomical survey. DESI has now accumulated multiple years of field experience, making it the most relevant performance benchmark for future systems.

The main drawback of the DESI architecture for future high-density focal planes is dimensional scaling. The DESI design uses small DC gear motors, but the motor diameter and associated mechanical structure make direct reduction to a pitch of 6.2\,mm difficult. Achieving such a pitch requires either smaller motors combined with a more compact mechanical transmission, or a fundamentally different actuation scheme.

\subsection{Compact Motorized Designs}

One possible solution to the scaling challenge is to retain motorized actuation while altering the mechanical integration. The Trillium design \cite{trillium}, for example, uses the same 4-mm brushless DC gear motors as DESI but introduces additional transfer gearing to couple the eccentric axis to the central rotation axis. This allows the motors to be nested axially rather than placed side by side, thereby reducing the effective pitch while retaining much of the DESI motor and electronics heritage.

Motorized designs benefit from operational heritage, high stiffness, and relatively straightforward control, as demonstrated by the mature DESI open-loop control system. However, they require precision gearing, bearings, and careful integration, while also providing sufficient torque to position the fiber. These issues become increasingly challenging as the pitch decreases.

\subsection{Piezoelectric Tilting-Spine Designs}

Tilting-spine architectures provide a different path to realizing a small pitch system. In systems such as Echidna\cite{Echidna1}, a fiber is mounted in a stiff spine that tilts to place the fiber tip on the desired target within its patrol region. The motion is generated by piezoelectric ``slip-stick actuation'': a sawtooth waveform drives a piezo element, and frictional coupling causes the spine to ``walk'' toward the desired location over many small steps. Echidna Mark II\cite{Echidna2} uses piezo stack actuators instead of tubular piezo elements as in earlier designs to improve precision and reduce voltage requirements.

Tilting spines scale naturally to small pitches because the actuator volume can be made compact and because the patrol region is generated by angular motion of a narrow spine. However, the slip-stick system depends on frictional interfaces, which are prone to variability and wear and therefore likely require a more elaborate calibration. In addition, any tilting-spine design has the fundamental consequence that the fiber tip moves along an arc rather than remaining exactly in the focal surface. This creates defocus and telecentricity errors that can contribute to FRD (focal-ratio degradation) unless mitigated. This effect becomes less problematic as the pitch decreases.  

\subsection{Compliant Flexure-Based Designs}

Another approach is to use a compliant mechanism for the radial positioning of the fiber. In such designs, elastic flexures provide radial motion through elastic deformation, producing smooth and repeatable displacement. 

One example is an R--$\theta$ architecture in which a conventional rotary stage provides azimuthal motion and a linear-motion flexure provides radial motion. In the R-FLEX concept\cite{Wenner2026}, small rotations of a cam at the base of the flexure are converted into a larger, nearly tilt-free radial displacement of the fiber tip. A compact motor drives the cam, which actuates thin leaf flexures arranged as a parallel linear spring.

Flexure-based designs are attractive for dense focal planes because they can be slim, have a low part count, and be compatible with mass production. Their geometry can be optimized for travel, stiffness, stress, defocus, and manufacturability. The main trade-offs are material stress, fatigue life and achievable travel. Overall, compliant R--$\theta$ mechanisms retain deterministic motor-driven actuation while replacing part of the mechanical transmission. The flexure itself is nearly backlash-free, and any backlash introduced into the system will stem from the gear motor driving the cam.

\section{Hybrid R--$\theta$ Positioner Concept}
\label{sec:concept}
\begin{figure} [ht]
\begin{center}
\begin{tabular}{c} 
\includegraphics[width=\linewidth]{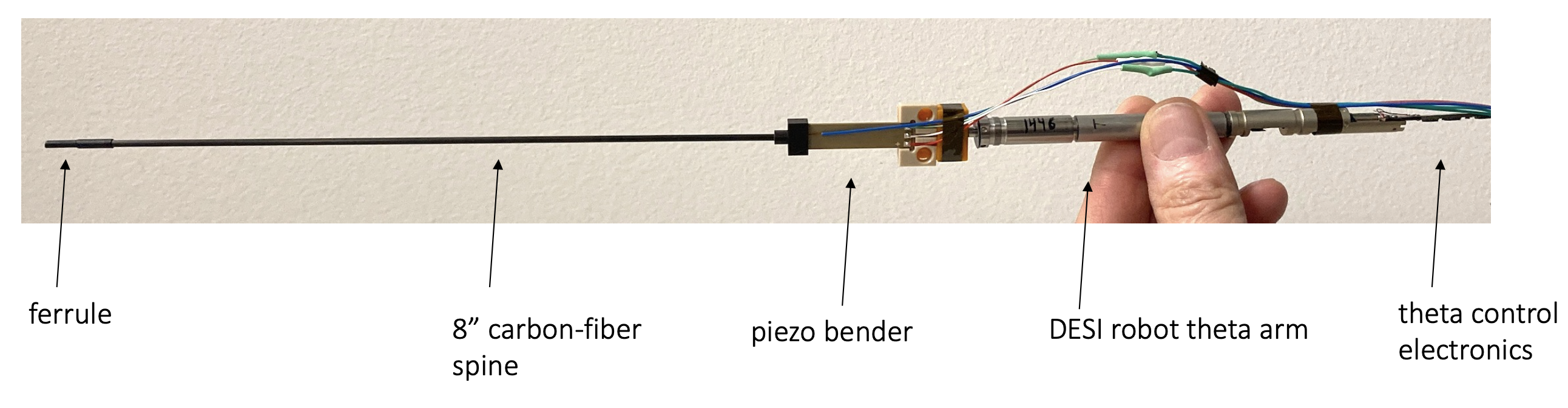}
\end{tabular}
\end{center}
\caption{Prototype hybrid R--$\theta$ positioner showing the DESI $\theta$ arm, piezoelectric bender, carbon-fiber spine, ferrule, and control electronics.}
\label{fig:prototype} 
\end{figure}

The hybrid design explored here combines a DESI-like $\theta$ motor with a piezoelectric bender for radial motion. In our prototype (see Fig.~\ref{fig:prototype}), rotation about the central axis is provided by the existing DESI motorized $\theta$ stage. This stage could be reduced in size, even with the current 4-mm motor, to match a 6.2-mm pitch design. The conventional $\phi$ arm is replaced by a piezoelectric bimorph bender coupled to an 8-inch carbon-fiber tube that carries the fiber ferrule at its tip. In this configuration, the $\theta$ motor controls the angular position, while the bender controls the radial displacement.

For the prototype, we used a Thorlabs PB4NB2S piezoelectric bimorph bender. This multilayer ceramic actuator deflects primarily along one axis and moves approximately along a circular arc with a radius comparable to the length of the bender. A simple custom 3-D-printed holder was fabricated to attach the 8-inch carbon-fiber tube to the bender. With this extension, the measured maximum deflection at the fiber end was approximately 6\,mm at 140\,V. Although the bender can, in principle, be driven over a bipolar range from $-D$ to $+D$, the prototype was operated over a unipolar range from $0$ to $D$. This simplifies the high-voltage drive requirements and allows the ferrule angle to be mechanically biased so that the effects of defocus and telecentricity errors are minimized near maximum deflection, where most of the target area lies.

The prototype was driven using a computer-controlled DAC providing a low-voltage input to a fixed-gain piezo amplifier. In our laboratory setup, a PiezoDrive BD300 amplifier was used to generate the high-voltage drive signal applied to the bender. A more compact commercial piezo driver may be suitable for future integrated positioner electronics, although voltage limits and channel count must be considered.

The bending plane of the piezo actuator must be clocked with respect to the $\theta$ rotation axis. Ideally, the bender deflects in the radial plane containing the $\theta$ axis, so that piezo motion changes only the radial coordinate. A clocking accuracy of order $0.5^\circ$ is sufficient and the residual uncertainty can be absorbed into the calibration and closed-loop feedback.

\section{Piezoelectric Behavior and Control}

Piezoelectric actuators exhibit an approximately linear relationship between applied voltage and displacement over limited ranges, with very fast intrinsic response. However, precision positioning is complicated by two well-known effects: creep and hysteresis.

Creep arises because the polarization domains in the piezoelectric material continue to reorient after a voltage step, causing the actuator displacement to evolve slowly toward a steady-state value. The creep behavior is often approximately logarithmic in time and can be modeled as

\begin{equation}
\Delta L(t) \approx \Delta L_{t=0.1\,\mathrm{s}}
\left[
1 + \gamma \log\left(\frac{t}{0.1\,\mathrm{s}}\right)
\right],
\label{eq:creep}
\end{equation}

where $\gamma$ is a material-dependent creep factor, typically of order 0.01--0.02.

Hysteresis causes the displacement-voltage relationship to depend on the previous voltage history. As voltage is increased and then decreased, the actuator follows different displacement paths. For open-loop positioning, this produces target-dependent and history-dependent errors. Therefore, precision motion with piezoelectric elements generally requires closed-loop control.

\begin{figure} [ht]
\begin{center}
\begin{tabular}{c} 
\includegraphics[width=10cm]{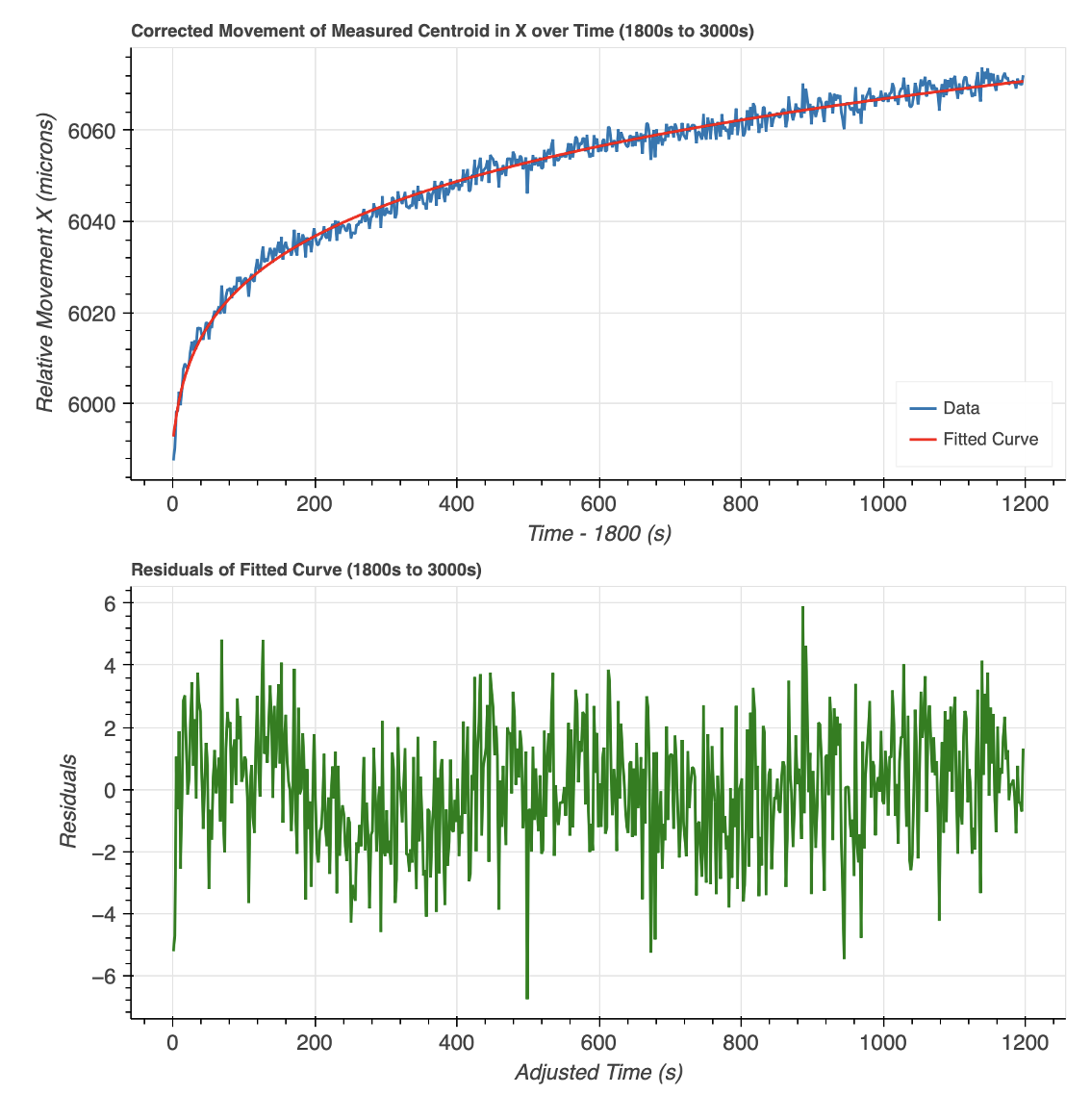}
\end{tabular}
\end{center}
\caption[example] 
{ \label{fig:fig2} 
Measured open-loop creep of the prototype bender after a voltage step, with logarithmic creep fit and residuals.}
\end{figure}

In our laboratory setup, feedback was provided using a fiber-view-camera measurement of the illuminated fiber position. FITS images from a CMOS camera were analyzed with a centroiding algorithm capable of operating at approximately 1 Hz. In the tests reported here, feedback corrections were applied every 2\,s.

The chosen closed-loop algorithm was intentionally simple. For a desired radial displacement, an initial drive voltage was estimated from a linear calibration. The voltage was then ramped up gently to the target value at approximately 10\,V per 0.1\,s to avoid exciting vibrations. The fiber position was then measured by the camera, a correction voltage was calculated from the measured centroid error, and the correction was applied. This process was repeated every 2\,s.

\section{Laboratory Performance}

Open-loop tests showed the expected piezoelectric creep (Fig.~\ref{fig:fig2}). Following a commanded displacement, the fiber position continued to drift over time, with behavior well described by the logarithmic creep model of Eq.~\ref{eq:creep}. This confirms that open-loop operation is not sufficient for micron-level fiber placement over relevant timescales( typically 15-20 minutes for galaxy spectra).

Closed-loop operation substantially reduced this drift. In a representative test, the fiber reached the commanded target after only two feedback corrections, corresponding to less than 5\,s total settling time. After convergence, the residual position error was approximately 2 $\mu$m RMS.

\begin{figure} [ht]
\begin{center}
\begin{tabular}{c} 
\includegraphics[height=5.5cm]{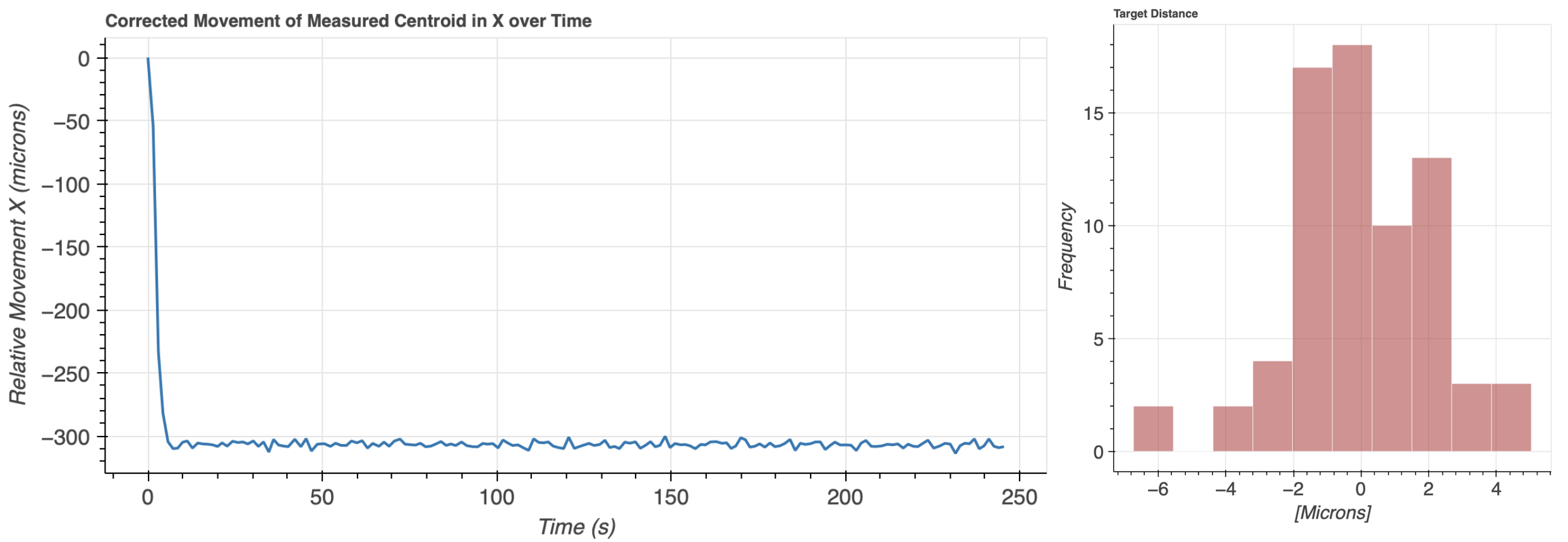}
\end{tabular}
\end{center}
\caption[example] 
{ \label{fig:example} 
Closed-loop response with feedback applied every 2 s. The fiber reaches the commanded target after two corrections and remains stable with approximately 2 $\mu$m RMS residual error.}
\end{figure}

A comparison between open-loop and closed-loop operation illustrates the central requirement for piezoelectric positioning. In open loop, the bender continues to drift after the initial voltage command. In closed loop, the camera-based feedback corrects the creep and maintains the fiber near the desired location. The result demonstrates that a piezo-bender radial actuator can achieve high precision, but only when integrated into an appropriate control architecture.

\section{Operational Considerations}

Several operational issues must be addressed before a piezo-bender R--$\theta$ positioner could be considered for a full focal-plane implementation.

First, because closed-loop operation is required, the system needs a reliable way to measure the fiber position during configuration. DESI uses back-illumination of the science fibers together with a fiber-view camera. For a system requiring repeated feedback during positioning, continuous or repeated back-illumination of the science fiber may not be practical in the same way. One possible solution is to mount a small illuminated fiber stub ('metrology fiber') adjacent to the science fiber. To avoid contamination of the science band, the metrology system could operate in the near-infrared; ultraviolet illumination may also be possible but is likely more challenging.

Second, the stiffness of the piezo-bender and extended carbon-fiber spine must be sufficient to survive handling, telescope motion, vibration, and environmental disturbances. Benders provide large displacement in a compact form but are less stiff than tubular actuators or motorized mechanical arms. This trade-off must be evaluated in the context of the required settling time, telescope orientation changes, and allowable fiber-position error.

Third, as with all tilting-spine-like designs, the fiber tip moves along an arc. It cannot remain exactly tangent to the focal surface over the full patrol region. As already discussed, this introduces defocus and telecentricity errors, which may increase focal-ratio degradation and reduce throughput. The severity of this effect is reduced for smaller-pitch positioners because the required patrol radius is smaller. Additional mitigation is possible through longer spines and biased ferrules, where the nominal un-deflected position is intentionally offset so that the most frequently used target positions occur at favorable tilt angles.

Finally, large-scale implementation would require compact high-voltage electronics, channel multiplexing or integrated drivers, thermal management, failure-mode analysis, and compatibility with a modular focal-plane architecture. These engineering issues are nontrivial but appear manageable given the small power requirements of capacitive piezoelectric actuation.

\section{Conclusions}

We have developed and tested a prototype hybrid R--$\theta$ robotic fiber positioner in which a DESI-like $\theta$ motor provides azimuthal rotation and a piezoelectric bimorph bender provides radial motion. The concept is motivated by the need for smaller-pitch fiber positioners in future massively multiplexed spectroscopic instruments, where direct scaling of DESI-like $\theta$--$\phi$ motorized positioners is challenging.

The prototype achieved approximately 6\,mm radial deflection at 140 V using a commercial bimorph bender and an 8-inch carbon-fiber spine. Open-loop measurements showed the expected piezoelectric creep and hysteresis, confirming that closed-loop operation is required for precision fiber placement. Using fiber-view camera centroid feedback applied every 2 s, the prototype reached its target after two corrections, in less than 5\,s, and maintained a residual stability of approximately 2 $\mu$m RMS.

Piezoelectric benders offer a compact means of generating large displacement at moderate voltage compared with tubular piezo actuators. Their primary disadvantages are reduced stiffness and the need for closed-loop correction of creep and hysteresis. In addition, any tilting-spine-like geometry introduces telecentricity and defocus errors because the fiber tip moves along an arc. These effects may be mitigated through long spines, biased ferrules, and careful optical design.

The results show that direct piezo-bender radial actuation is a promising design direction for dense focal-plane fiber positioners. Future work will focus on integrated compact high-voltage electronics, improved mechanical stiffness, detailed telecentricity and focal-ratio-degradation modeling, multi-positioner packaging, and closed-loop metrology schemes compatible with large focal-plane operation.

\appendix    

\acknowledgments 
 
This work was supported in part by the Director, Office of Science, Office of High Energy Physics of the US Department of Energy under contract No. DE-AC02-05CH11231. The authors acknowledge the contributions of the broader Spec-S5 collaboration and the many institutions supporting the development of next-generation spectroscopic survey facilities. This work builds upon the heritage of SDSS, BOSS, eBOSS, DES, DESI, and related survey projects."

\bibliography{report} 
\bibliographystyle{spiebib} 

\end{document}